\documentclass[preprint1]{aastex}
\newcommand{\bfv}{\mbox{\boldmath $v$}}
\newcommand{\bfu}{\mbox{\boldmath $u$}}

\newcommand{\bnabla}{\mbox{\boldmath $\nabla$}}
\newcommand{\radvel}{V_{\parallel}}
\slugcomment{submitted to ApJL}
\shorttitle{Pattern speed of the nucleus of M31}
\shortauthors{Sambhus and Sridhar}

\begin{document}
\title{The pattern speed of the nuclear disk of M31\\
using a variant of the Tremaine--Weinberg method}
\author{Niranjan Sambhus and S. Sridhar}
\affil{Inter--University Centre for Astronomy and Astrophysics\\
Ganeshkhind, Pune 411007, India}
\email{nbs@iucaa.ernet.in\\
sridhar@iucaa.ernet.in}

\begin{abstract}
The twin peaks in the nucleus of M31 have been interpreted by Tremaine
as a thick, eccentric, disk of stars orbiting a massive dark object;
the required alignment of the apoapsides of the stellar orbits could
be maintained by self--gravity, and the whole structure might 
be a discrete, nonlinear eigenmode. The pattern speed of this
mode could, in principle, be determined by the Tremaine--Weinberg (TW) 
method, which requires measurements of the surface brightness, and 
radial velocity along a strip parallel to the line of nodes. 
However, spectroscopic observations along the line of nodes are not
available.  We propose a variant of the TW method, which
exploits a basic feature of  the eccentric disk model, to extract 
estimates of the pattern speed from {\it Hubble  Space Telescope}
spectroscopic data, taken along the line joining the two peaks. 
Within limitations imposed by the data, we estimate that the 
pattern rotates in a prograde manner and, for an assumed disk inclination of 
$77^{\circ}$, the pattern speed $\left|\Omega_p\right| < 30\;{\rm 
km\,s^{-1}\,pc^{-1}}$, or period more than $200,000$~years. 
\end{abstract}  


\section{Introduction}
The nucleus of M31 was first resolved by the Stratoscope~II 
balloon--borne telescope (Light, Danielson, and Schwarzschild~1974), 
which showed that the peak brightness was displaced relative to the 
center, as inferred from the isophotes of the outer parts of the galaxy.
This was confirmed, and extended by {\it Hubble Space Telescope} 
({\it HST}) observations, which  revealed two peaks in the 
brightness, separated by $0''.49\,$~(Lauer et al.~1993, King,
Stanford, and Crane~1995). Ground--based, as well as {\it HST}
spectroscopy, have probed the structure of the radial velocities 
and velocity dispersions, in increasing detail, along many strips across
the nuclear region (Dressler and Richstone~1988, 
Kormendy~1988, Bacon et al.~1994, van der Marel et al.~1994, 
Gerssen, Kuijken, and Merrifield~1995, Statler et al.~1999, 
Kormendy and Bender~1999). These provide evidence for the presence of a 
massive dark object (MDO), which  could be a supermassive black  hole, 
of mass $\sim 3\times  10^7\,{\rm\,M_\odot}\,$, located very close to the 
fainter peak (P2). The dynamical center of the nucleus is believed to 
coincide with the center of the isophotes of the bulge of M31; this 
point has been estimated to lie between the two peaks.  

Tremaine~(1995) proposed that the nucleus could be a thick
eccentric disk, composed of stars on nearly Keplerian orbits around the
MDO, with their  apoapsides aligned in the direction toward the brighter
peak  (P1); the brightness of P1 is then explained as the increased
concentration of  stars, resulting from their slow speeds near their
apoapsides. He also showed that this model is broadly consistent
with the kinematics, as inferred from the spectroscopic observations
of Kormendy~(1988), and Bacon et al.~(1994).  
Recent work has not only produced further support for Tremaine's 
model (Statler et al.~1999, Kormendy and Bender~1999), but has 
stimulated variations on the basic theme (Statler~1999). Tremaine also
suggested that the alignment could be maintained by the
self--gravity  of the disk, wherein the eccentric distortion could arise
as a discrete, nonlinear eigenmode, with some nonzero pattern speed
($\Omega_p$), equal to the common  apsidal precession rate. The dynamical
question  is yet to be resolved in a self--consistent manner, although 
explorations of orbits in model potentials have identified a family of 
resonant, aligned loop orbits, which could serve as building blocks
(Sridhar and Touma~1999, Statler~1999); reasonably faithful reproduction
of the nuclear rotation curve adds some degree of confidence 
(Statler~1999). If the nuclear disk is indeed a steadily rotating, 
nonlinear eigenmode, what is $\Omega_p\,$? 
  
Tremaine and Weinberg~(1984, hereafter TW) invented a method of 
estimating the pattern speed of a barred disk galaxy, that
uses measurements  of the surface brightness, and radial velocity along        
a strip parallel  to the line of nodes (defined as the line of 
intersection of the disk  and sky planes). That this methods works was 
proved when the pattern speed of the bar in NGC~936 was estimated by 
Merrifield and Kuijken~(1995). Unfortunately, the radial velocity
measurements of the nucleus of M31 (see references above on spectroscopy)     
are available on strips that, either do not coincide with the line of
nodes, or possess too poor an angular resolution for direct application
of  the TW method. In this Letter, we show that the {\it HST} 
observations of Statler et al.~(1999, hereafter SKCJ), together with 
Tremaine's (1995) model, can be used to estimate $\Omega_p\,$.

\section{A modified Tremaine--Weinberg method applicable to M31}
We briefly recall TW's derivation of their kinematic method.  
Let us assume that the disk is razor--thin, with a well--defined 
pattern speed, $\Omega_p\,$. The plane of the disk is assumed to 
be inclined at angle $i$ with respect to the plane of the sky.
Let $(x, y)$, and $(r, \phi)$ be 
cartesian and polar coordinates, respectively, in the plane of the 
disk, with the origin coinciding with the center of mass of the 
system (disk plus MDO). Let the cartesian coordinates in the sky 
plane be $(X, Y) = (x, \,y\cos{i})\,$; the $x$ and $X$ axis coincide 
with the line of nodes. The disk is assumed to rotate steadily, hence 
the surface brightness of stars, $\Sigma(x, y, t) =  \tilde{\Sigma}
(r, \phi - \Omega_pt)\,$. The surface brightness is  also assumed 
to obey a continuity equation, without a source term. Let $\bfv$ be the 
(mean) velocity field in the disk plane. The continuity equation can 
be manipulated to yield, 
\begin{equation}
\bnabla\cdot\left[\left(\bfv - \bfu\right)\Sigma\right] \;=\; 0\,, 
\label{contin}
\end{equation}
\noindent where $\bfu = \Omega_p(-y\hat{x} + x\hat{y})\,$. 
TW proceed by integrating  equation~(\ref{contin}) over $x$ 
from $-\infty$ to $+\infty$. Assuming $\Sigma\to 0$ sufficiently rapidly
as $|x|\to \infty$,  and integrating over $y$ from $y$ to $\infty$ yields
$\Omega_p\int_{-\infty}^{\infty}dx\,\Sigma\,x = \int_{-\infty}^{\infty}
dx\,\Sigma\,v_y\,$. Noting that the sky brightness, $\Sigma_s =
\Sigma/\cos{i}\,$, and the radial velocity, $\radvel = v_y\sin{i}$,
the integrals may be expressed in terms of observable quantities;
hence $\Omega_p$ can be estimated when $\radvel$ has been measured 
on a strip parallel to the line of nodes.  

SKCJ measured $\radvel$ along the P1--P2 line,
which is inclined by about $4^{\circ}$, in the sky plane, to the line 
of nodes. Therefore $\radvel$ is available on a strip that makes an
angle, $\theta\simeq 4^{\circ}/\cos 77^{\circ}\simeq 18^{\circ}\,$, 
with the $x$--axis, in the disk plane; the TW procedure needs some 
modification, before $\Omega_p$ can be extracted. Let $x' = x\cos\theta 
+ y\sin\theta\,$ and $y' =  -x\sin\theta + y\cos\theta\,$ be the rotated
cartesian coordinates. Let us also denote the surface brightness by
$\Sigma'(x', y', t)\,$. The SKCJ measurements are 
along the strip $y' = 0\,$, that passes through the origin.
Equation~(\ref{contin}) expresses a relation that is invariant under
rotation of cartesian coordinates. Hence application of the TW procedure 
provides an identical relationship between the integrals over the strip,
defined by $y' = 0\,$: 
\begin{equation}
\Omega_p\int_{-\infty}^{\infty}dx'\,\Sigma'\,x' \;=\; 
\int_{-\infty}^{\infty}dx'\,\Sigma'\,v_y'\,
\label{relation}
\end{equation}
\noindent We now express the integrals in terms of observable 
quantities: $x' = X\cos\theta + Y\sin\theta/\cos{i}\,$ and $0 = y' =
-X\sin\theta + Y\cos\theta/\cos{i}\,$ can be used to 
eliminate $Y$, giving $x' = X/\cos\theta\,$. As before, 
$\Sigma'(x', y', t) = \cos{i}\,\Sigma_s(X, Y, t)\,$. However, the radial 
velocity, 
\begin{equation}
\radvel = \sin{i}\left(v_y'\cos\theta + v_x'
\sin\theta\right)\,,
\label{radvel}
\end{equation}
\noindent depends on $v_y'$ as well as $v_x'\,$. Hence it 
is, in general, not possible to express $v_y'$ in 
equation~(\ref{relation}) in terms of $\radvel$ alone (the TW 
method finesses this problem because $\theta = 0$ kills the 
contribution to $\radvel$ from the $v_x'$ term). However, we are able 
to make progress by 
recalling the basic elements of Tremaine's (1995) model. {\it If} the 
nuclear  disk is largely composed of nearly Keplerian ellipses, with 
their apoapsides aligned along the P1--P2 line, {\it and} we choose 
$\theta$ such that our strip lies along this line, then the strip 
intersects all these ellipses at right angles. Thus $v_x' = 0\,$, 
and we recover the familiar TW relation,  \begin{equation}
\Omega_p\,\sin{i}\int_{-\infty}^{\infty}\,dX\,\Sigma_s\,X \;=\;
\int_{-\infty}^{\infty}\,dX\,\Sigma_s\,\radvel\,.
\label{twrelation}
\end{equation}
\noindent It should be noted that the value of $\theta$ drops out
of equation~(\ref{twrelation}). Moreover, $v_x' =0$ even when the 
apsides of the ellipses precess. Below we use {\it HST} photometry
and kinematics to estimate $\Omega_p\,$. 
         
\section{$\Omega_p$ from {\it HST} photometry and spectroscopy}
{\it HST} photometric data, reported in Lauer et al.~(1993), was kindly
supplied to us by Prof.~Ivan~King. Figure~1 shows the (sky) surface
brightness along the P1--P2 line. The contribution from 
the bulge of M31 was subtracted using two different fits to the 
bulge brightness, namely a Nuker fit as described in Tremaine~(1995), 
and a S\'ersic fit as described in Kormendy and Bender~(1999). The 
disks so obtained will henceforth be referred to as a ``T--disk'', 
and a ``KB--disk'', respectively.

\begin{figure}[t]
\plotone{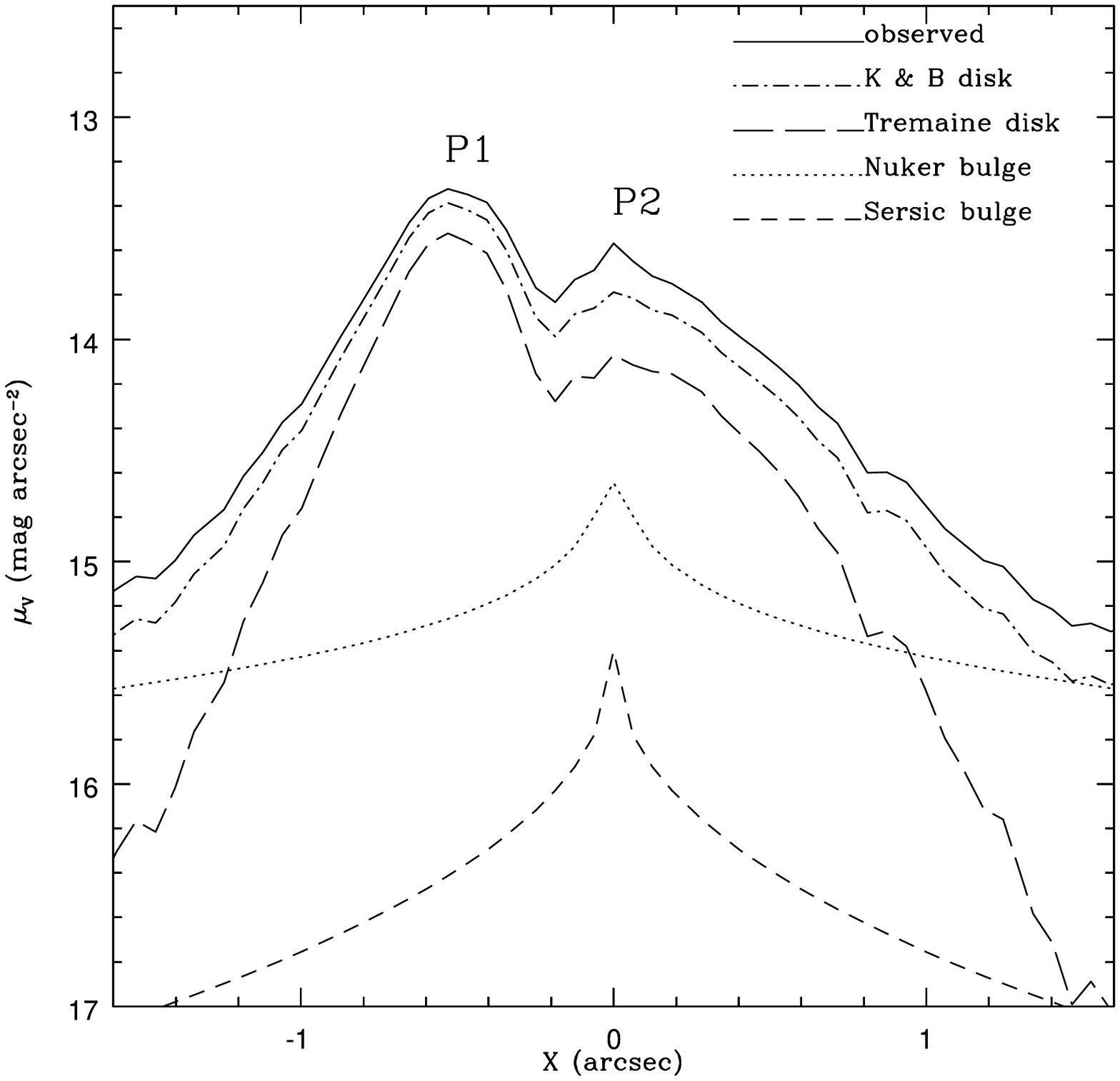}
\figcaption[photometry.ps]{V--band surface brightness along
the P1--P2 line, with P2 located at $X=0$, and P1 at $X= -0''.49$.
The solid line connects data points taken from Lauer et al.~(1993).
The dotted and short--dashed curves are the contributions from
Nuker (Tremaine~1995) and S\'ersic (Kormendy and Bender~1999) bulges,
respectively. The corresponding estimates for the surface brightness
of the disk (T--disk and KB--disk) are given by the long--dashed, and
dot--dashed curves.
\label{fig1}}
\end{figure}

\begin{figure}[t]
\plotone{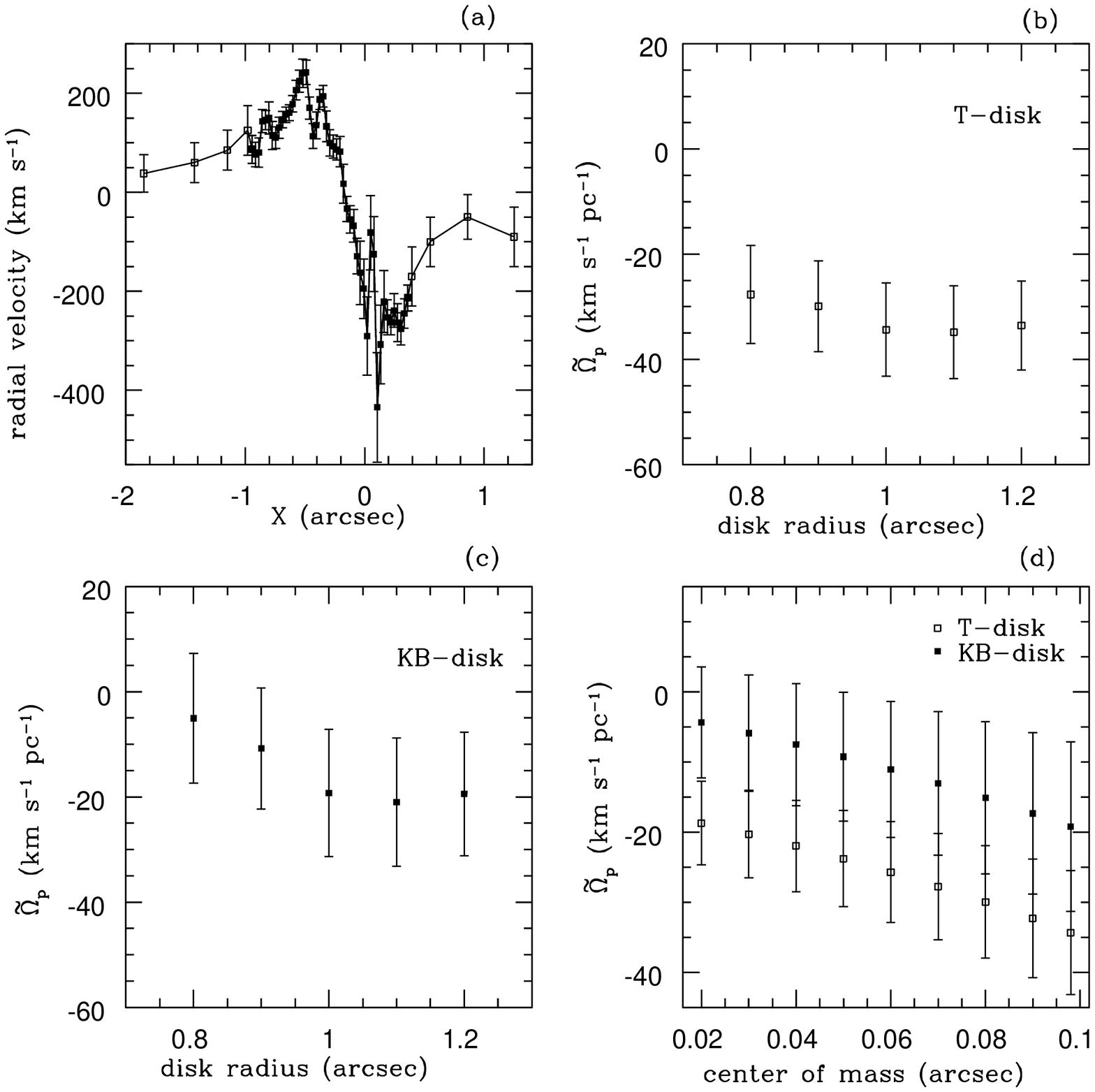}
\figcaption[kinematics.ps]{(a) Radial velocity measurements
taken from SKCJ. (b, c) $\tilde{\Omega}_p$ plotted versus the truncation
radius of the disk, as measured from the center of mass of the
system, for the T--disk and KB--disk, respectively; the center of mass
was taken to be $0''.098$ away from P2 toward P1.
(d) $\tilde{\Omega}_p$ plotted versus the assumed location of the center of
mass of the system, for a disk truncation radius equal to $1''.0$.
\label{fig2}}
\end{figure}       

SKCJ observed the stellar kinematics along the P1--P2 
line, using the $f/48$ long--slit spectrograph of the {\it HST} Faint 
Object Camera. We obtained radial velocities along the P1--P2 line, 
including the errors on them,  from their ``de--zoomed'' 
rotation curve (given in Figure~3, as well as Table~1 of 
SKCJ); these are displayed in Figure~2a. The integrals in 
equation~(\ref{twrelation}) need to be computed, with upper and lower limits
symmetrically displaced about the center of mass of the disk plus MDO.
Kormendy and Bender~(1999) have determined the center of mass to be
displaced by $0''.098$ from P2 toward P1, and we use this value in 
the computations for Figures~2b and 2c. The stated errors on the radial 
velocities were used by us to generate 300 random  realizations. For each 
of these realizations of the rotation curve, we evaluate the integrals for 
five different limits, ranging from  $\pm 0''.8$ to $\pm 1''.2\,$. 
As is clear from equation~(\ref{twrelation}), only the combination, 
$\Omega_p\,\sin i$ can be determined. It is a common assumption that the
nuclear disk of M31 has the same inclination, to the sky plane, as the
much larger galactic disk of M31, which is inclined at $\simeq 77^{\circ}\,$.
We wish to state our results independent of this assumption, so we
present estimates of the quantity, $\tilde{\Omega}_p = 
(\sin i/\sin 77^{\circ})\,\Omega_p\,$, in Figures~2b---2d.

Figures~2b and 2c display the estimates of  $\tilde{\Omega}_p$ 
so obtained, together with $1\sigma$ error bars, as a function of the limits
of the integrals, for the T--disk and KB--disk, respectively. It is evident
that the estimates of pattern speed do not vary much when the limits of
integration lie between $\pm 1''.0$ and $\pm 1''.2\,$. We also explore the
dependence on the position of the center of mass of the system, assumed in the
computation of the integrals. Although we used the most recent determination,
due to Kormendy and Bender~(1999), it must be noted that earlier work
(Lauer et. al. 1993, King, Stanford, and Crane~1995, Tremaine~1995)
records smaller values, $\leq 0''.05$, away from P2 toward P1. Figure~2d
plots our estimates of $\tilde{\Omega}_p$, for a range of values of the center
of mass, with limits of integration fixed at $\pm 1''.0\,$. A negative value
of $\tilde{\Omega}_p$ means that the pattern is {\it prograde};
Figures~2b---2d indicate that this is the most likely possibility, with the
absolute value increasing with the separation between P2 and the center of
mass of the system. This is reasonable, because a larger separation
corresponds to a larger disk mass, which  implies a greater contribution from
the self--gravity of the disk, which  is ultimately responsible for the
(alignment and) precession of the disk. The errors on $\tilde{\Omega}_p$ are
large, and a non rotating pattern cannot be  ruled out with certainty. Below
we quote representative  bounds on the absolute value of the prograde pattern
speed:  
\begin{equation}
\left|\Omega_p\right|  \leq \frac{\sin 77^{\circ}}{\sin i}
\cases{34 \pm 8\;{\rm km\,s^{-1}\,pc^{-1}},
&T--disk;\cr 		  20 \pm 12\;{\rm km\,s^{-1}\,pc^{-1}},
&KB--disk.\cr} 
\label{pattern}
\end{equation}
\noindent For each realization of the rotation curve, the
zero--crossing  point (the ``rotation center'') was determined by a
third--order spline interpolation. We found the position of the 
rotation center to be displaced by $0''.17\pm 0''.01$, from P2 toward P1.
This should be compared with the value of $0''.16 \pm 0''.05$, quoted 
by SKCJ. We also tested for any systematic dependence of $\Omega_p$
on the position of the rotation center, and found none. 

\section{Conclusions}
Our estimate of the pattern speed of the nuclear disk of M31
should be qualified by a discussion of possible sources
of errors, most of which are difficult to estimate 
quantitatively. SKCJ calibrate velocities relative to an 
average over an $8\,{\rm arcsec}^2$ aperture centered on the nucleus
(Ho, Filippenko, and Sargent~1993), and assure us that the errors
are likely to be small. More significant, perhaps, are the systematic 
errors in $\radvel$, mentioned by SKCJ; these are shown as open squares in
Figure~2a. SKCJ used a slit of width $0''.063$, and 
this will introduce contributions to $\radvel$ from nonzero values of
$v_x'$ (see  equation~(\ref{radvel})). This effect is somewhat mitigated by
cancellation  between positive and negative values of $v_x'$, and the 
fact that the width of $0''.063$ is of much smaller scale than the 
minor axis, projected onto the sky plane, of the smallest ellipse
used by Tremaine~(1995; see Figure~2a of his paper) to represent 
the nuclear disk. The limits of integration are necessarily
finite in numerical computation, and we have resisted the temptation to
include corrections by extrapolation of the $\Sigma_s$ and $\radvel$
profiles.   

A basic assumption underlying the application of a TW--like method is 
that the surface brightness obeys a continuity equation, which would be 
valid for a stellar disk in the absence of star formation (or death).
We expect the numbers of stars to be conserved, except possibly
in the vicinity of P2, where the observed UV excess has been 
interpreted as contributions from early--type stars (King, Stanford and 
Crane~1995, Lauer et al.~1998). However, these stars do 
not contribute much to the photometric and kinematic data we have 
used. Tremaine~(1995) argues that two--body relaxation is expected to 
thicken the disk, whereas we assumed that the disk was razor--thin.
The original TW method is applicable to thick disks, 
so long as the streaming velocity normal to the disk plane is zero.
In addition to the assumption of zero normal streaming velocity, let 
us suppose that the three dimensional density, $\rho$, is symmetric about 
the midplane of the disk. The contribution to $\radvel$ from $v_x'$ arise 
from an integral along the (inclined) line of sight that runs through the 
thick disk.  Consider two points along this line of sight that are equally
displaced about the midplane of the disk. Aligned, nearly Keplerian orbits
have  flows such that $\rho$ is equal, whereas $v_x'$ is equal and opposite
at these two points; in this ideal picture, there is pair--wise perfect
cancellation, and no net contribution to $\radvel$ from $v_x'\,$. In 
practice there should be some cancellation, because $v_x'$ will have 
opposite signs at two corresponding points, but there could be a net 
contamination from the unequal values of $|v_x'|$ and $\rho$\,.

Tremaine's original model, which was a reasonable fit to the then available
photometry and kinematics, considered a non rotating disk, and it would 
be appropriate to inquire about the implications of a non zero pattern speed.
A pattern that is prograde with angular speed, say, of $20\;{\rm
km\,s^{-1}\, pc^{-1}}$ would contribute only about $35\;{\rm km\,s^{-1}}$ to
the radial  velocity at P1, which is about $250\;{\rm km\,s^{-1}}\,$,
according to the measurements of SKCJ. The maximum radial velocity quoted by 
Tremaine~(1995) is less than $200\;{\rm km\,s^{-1}}\,$, so a non zero pattern 
speed could still be accomodated. Our estimates do not rule out a non 
rotating disk, but we would like to offer a physical argument in support
of a non zero pattern speed. For the disk plus MDO to be in a steady, non 
rotating state, the gravitational force on the MDO should necessarily 
vanish. Our (unpublished) numerical computations indicate that the 
force is indeed non zero. 

A limitation of our method is that it uses, in an essential manner, the 
assumption that most of the contribution to $\radvel$ comes from orbits 
which intersect the measurement strip at right angles. In comparison, the 
original TW method does not rely on assumptions about the geometry of the 
mean flow; averaging over several strips, all parallel to the line of nodes, 
will improve estimates of the pattern speed, as Merrifield and Kuijken~(1995)
demonstrated. Thus it is necessary to verify our  estimates of $\Omega_p$,
by using the TW method on future observations of  $\radvel$, together with
better photometry such as Lauer et al.~(1998), along strips parallel to the
line of nodes. An extremely useful set of observations that could be 
performed would be two--dimensional spectroscopy, similar to the work 
of Bacon et al.~(1994), with the increased angular resolution that should 
be available in the near future.  

\acknowledgments

We are grateful to Prof.~Ivan King for generously sharing with us 
the photometric data, an anonymous referee for pointing out a basic error 
in the original manuscript, and asking stimulating questions, and to
R.~Srianand for  useful comments. NS thanks the Council of Scientific
and Industrial Research, India, for financial support through grant 
2--21/95(II)/E.U.II.


\begin{thebibliography}{}
\bibitem[]{} Bacon, R., Emsellem, E., Monnet, G., and Nieto, J.L.  1994,
\aap, 281, 691
\bibitem[]{} Dressler, A., and Richstone, D. O., 1988, \apj, 324, 701
\bibitem[]{} Gerssen, J., Kuijken, K., and Merrifield, M. R. 1995, 
\mnras, 277, L21
\bibitem[]{} Ho, L. C., Filippenko, A. V., and Sargent, W. L. W. 1993, 
\apj, 417, 63
\bibitem[]{} King, I. R., Stanford, S. A., and Crane, P. 1995, \aj, 
109, 164
\bibitem[]{} Kormendy, J. 1988, \apj, 325, 128
\bibitem[]{} Kormendy, J., and Bender, R., 1999, \apj, 522, 772
\bibitem[]{} Lauer, T. R., {\it et al}. 1993, \aj, 106, 1436
\bibitem[]{} Lauer, T. R., {\it et al}. 1998, \aj, 116, 2263
\bibitem[]{} Light, E. S., Danielson, R. E., and Schwarzschild, M.
1974, \apj, 194, 257
\bibitem[]{} Merrifield, M. R. and Kuijken, K. 1995, \mnras, 274, 933
\bibitem[]{} Sridhar, S., and Touma, J. 1999, \mnras, 303, 483
\bibitem[]{} Statler, T. S., King, I. R., Crane, P., and 
Jedrzejewski, R. I. 1999, \aj, 117, 894
\bibitem[]{} Statler, T. S. 1999, \apjl, 524, L87
\bibitem[]{} Tremaine, S. and Weinberg, M. D. 1984, \apjl, 282, L5
\bibitem[]{} Tremaine, S. 1995, \aj, 110, 628
\bibitem[]{} van der Marel, R. P. {\it et al}. 1994, \mnras, 268, 521 
\end{thebibliography}
\end{document}